\documentclass[authoryear,preprint,review,12pt]{elsarticle}

\usepackage{amssymb}
\usepackage{amsmath}
\usepackage{mhchem}
\usepackage{soul,color}

\usepackage{lineno}

\journal{Icarus}

\newcommand{\about}{\mathop{\sim}\!}
\newcommand{\barstar}[1]{\Bar{#1}^{*}}
\newcommand{\Ro}   {\mathrm{Ro}}
\renewcommand{\vec}[1]{\mathbf{#1}}

\begin{document}

\begin{frontmatter}



\title{Reconciling Jupiter’s Vertical Motions with the Observed Cloud Structure in the Upper Troposphere}

\author[add1,add2,add3,add4]{Jo\~ao M. Mendon\c ca}
\author[add2]{Tapio Schneider}
\author[add2,add5]{Junjun Liu}
\author[add6]{Yuan Lian}
\cortext[1]{Corresponding authors emails: j.mendonca@soton.ac.uk, tapio@caltech.edu} 

\address[add1]{National Space Institute, Technical University of Denmark, Elektrovej, 2800 Kgs. Lyngby, Denmark}
\address[add2]{California Institute of Technology, Pasadena, CA 91125, USA}
\address[add3]{Department of Physics and Astronomy, University of Southampton, Highfield, Southampton SO17 1BJ, UK}
\address[add4]{School of Ocean and Earth Science, University of Southampton, Southampton, SO14 3ZH, UK}
\address[add5]{Space Science Institute, Boulder, CO 80301, USA}
\address[add6]{Aeolis Research, Chandler, AZ, 85224, USA}

\begin{abstract}
The eddy fluxes of angular momentum in Jupiter's upper troposphere are known to converge in prograde jets and diverge in retrograde jets. Away from the equator, this implies convergence of the Eulerian mean meridional flow in zones (anticyclonic shear) and divergence in belts (cyclonic shear). It indicates lower-tropospheric downwelling in zones and upwelling in belts because the mean meridional circulation almost certainly closes at depth. Yet the observed banded structure of Jupiter's clouds and hazes suggests that there is upwelling in the brighter zones and downwelling in the darker belts. Here, we show that this apparent contradiction can be resolved by considering not the Eulerian but the transformed Eulerian mean circulation, which includes a Stokes drift owing to eddies and is a better approximation of the Lagrangian mean transport of tracers such as ammonia. The potential vorticity structure inferred from observations paired with mixing length arguments suggests that there is transformed Eulerian mean upwelling in zones and downwelling in belts. Simulations with a global circulation model of Jupiter's upper atmosphere demonstrate the plausibility of these inferences and allow us to speculate on the band structure at deeper levels.
\end{abstract}


\end{frontmatter}

\section{Introduction}
\begin{figure*}[htb]
\centerline{\includegraphics{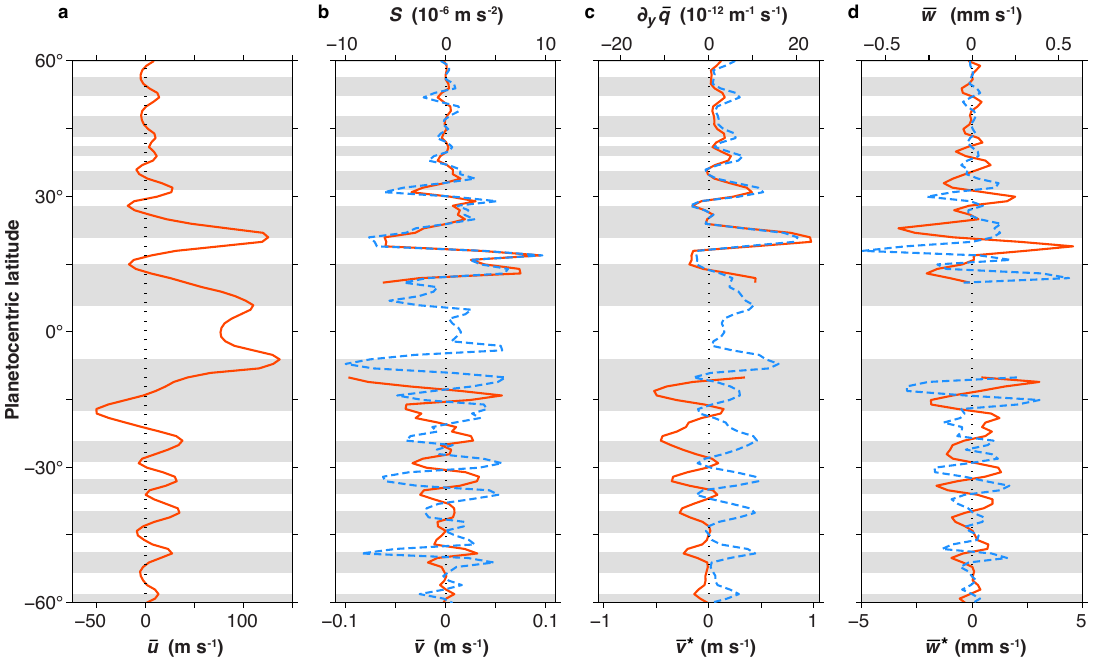}}
\caption{Zonal, meridional, and vertical velocities in Jupiter's upper troposphere. (a) Mean zonal velocity inferred from Cassini observations \citep{Salyk06}. (b) Eulerian mean meridional velocity (orange, lower axis) estimated as $\Bar v \approx S/f$ from the eddy momentum flux divergence $S$ (dashed blue, upper axis) inferred from Cassini observations \citep{Salyk06}. (c) Transformed Eulerian mean meridional velocity (orange, lower axis) estimated as  $\barstar{v} \approx (K/f) \partial_y \Bar q$ ($K=6\times10^6\,\mathrm{m\,s^{-2}}$) from the approximate potential vorticity gradient $\partial_y \Bar q \approx \partial_y (f + \Bar\zeta)$ (dashed blue, upper axis) implied by the zonal velocity in (a). (d) Eulerian mean (blue dashed, upper axis) and transformed Eulerian mean (orange, lower axis) vertical velocities estimated as $(\Bar w, ~\barstar{w}) = (H/2)\, \mathrm{div}(\Bar v,~\barstar{v})$. Estimated velocities are not shown within $\pm 10^\circ$ latitude, where the underlying approximations become inaccurate. Grey-shaded regions mark belts (cyclonic shear zones), in which the structure of clouds and hazes indicates downwelling; white regions mark zones, in which the structure of clouds and hazes indicates upwelling.}
\label{f:obs}
\end{figure*}

Observations of the structure of clouds and hazes and of turbulent momentum fluxes lead to apparently conflicting accounts of vertical motion in Jupiter's upper troposphere. On the one hand, the banded structure of Jupiter's clouds and hazes suggests that there is upwelling in the brighter zones and downwelling in the darker belts. (Zones are latitude bands of anticyclonic meridional shear of the zonal wind, and belts are latitude bands of cyclonic shear; see Fig.~\ref{f:obs}a). Upwelling in the zones is thought to lead to condensation of ammonia into relatively bright ice clouds in the upper troposphere; downwelling in the belts is thought to suppress ammonia ice cloud formation, revealing darker clouds and hazes at greater depth \citep[e.g.,][]{Smith79a,Gierasch86a,Carlson94a,Porco03,West04a,fletcher21}.

On the other hand, momentum fluxes associated with large-scale ($\gtrsim 1000$~km) eddies suggest the opposite sense of vertical motion in zones and belts. Eddy momentum fluxes in the upper troposphere generally converge in prograde zonal jets and diverge in retrograde jets \citep{Ingersoll81,Salyk06}; that is, they converge near the equatorward edges of the belts and diverge near their poleward edges (Fig.~\ref{f:obs}b, dashed blue line). But outside the equatorial region, where the Rossby number is small, the dominant balance in the zonal momentum equation \citep{Schneider09a} is 
\begin{equation}\label{e:zonmom0}
	f  \Bar v \approx S;
\end{equation}
that is, the Coriolis acceleration (Coriolis parameter $f$) of the Eulerian mean meridional flow $\Bar v$ balances the horizontal eddy (angular) momentum flux divergence $S$ to leading order, where
\begin{equation}\label{e:emfd}
S = \frac{1}{a \cos^{2}\phi} \, \frac{\partial}{\partial\phi} (\overline{u'v'} \cos^{2}\phi),
\end{equation}
$u$ and $v$ are the zonal and meridional velocities, and $a$ the planet's radius. Overbars denote a temporal and zonal mean at constant pressure; primes denote deviations therefrom. (Vertical eddy momentum fluxes do not enter to leading order because the horizontal flow is approximately geostrophic and non-divergent \citep[e.g.,][]{Andrews87b}. Hence, the eddy momentum flux convergence ($S < 0$) responsible for the prograde acceleration in the upper troposphere implies equatorward Eulerian mean meridional flow $\Bar v$; the eddy momentum flux divergence ($S > 0$) responsible for the retrograde acceleration implies poleward Eulerian mean meridional flow $\Bar v$ (Fig.~\ref{f:obs}b, orange line). This leads to convergence of the Eulerian mean meridional flow in zones and divergence in belts. For mass to be conserved, this upper-tropospheric meridional flow must be part of a larger overturning circulation. Energetic constraints require that the eddy momentum fluxes have a strongly baroclinic structure and are confined to a shallow layer in the upper troposphere, as they are on Earth \citep{Ait-Chaalal15a}; if they extended unabatedly over great depths, the implied conversion rate from eddy to mean-flow kinetic energy would exceed Jupiter's total atmospheric energy uptake \citep{Schneider09a,Liu10a,Thomson16a}. This means the \emph{meridional} branches of the Eulerian mean circulation are also confined to this shallow upper layer. The circulation must still close at depth, and it does so via its \emph{vertical} branches, linking the upper troposphere to a deep layer where drag acts on the flow \citep{Haynes91}. This principle of ``downward control'' suggests lower-tropospheric downwelling in zones and upwelling in belts (Fig.~\ref{f:obs}d, dashed blue line). Thus, there exists an apparent contradiction between inferences drawn from dynamical observations and from the structure of clouds and hazes.

Here, we show that this apparent contradiction can be resolved if one does not consider the Eulerian mean flow but an approximately Lagrangian mean flow. In the mean, tracers in turbulent flows are generally not advected by the Eulerian mean flow but by a Lagrangian mean flow, which contains a contribution owing to Stokes drift. \citet{Andrews76a,Andrews78a} showed that the Lagrangian mean flow advecting nearly conserved tracers can be approximated by adding an eddy contribution representing the Stokes drift to the Eulerian mean flow. We will show with scaling arguments and with numerical simulations that once the Stokes drift is taken into account, observations of Jupiter's tropospheric dynamics are consistent with the upwelling in zones and downwelling in belts suggested by the structure of clouds and hazes. In the resulting picture, the vertical motion is primarily accomplished by turbulent eddies, not the Eulerian mean flow. Since our study is based on first-order scaling arguments applied to data with measurement uncertainties, we validate the plausibility of our results with simulations from a 3D general circulation model of Jupiter's troposphere, which self-consistently reproduces the relevant atmospheric dynamics.


\section{Scaling arguments based on zonal momentum balance}\label{s:scaling}

\subsection{Eulerian mean flow}

The zonal momentum (or angular momentum) balance in Jupiter's troposphere is relatively simple because the Rossby number $\Ro = U/(fL)$ is small. With horizontal velocity scale $U \sim 10\,\mathrm{m\,s^{-1}}$ (appropriate outside the equatorial region, see \citealp{Salyk06}), scale of horizontal flow variations $L\sim 2000\,\mathrm{km}$ (Fig.~\ref{f:obs}a), and Coriolis parameter $f=2\Omega\sin\phi$ (planetary rotation rate $\Omega=1.76\times10^{-4}$ s$^{-1}$, \citealp{Donivan69}), we have usually $\Ro < 0.2$ outside $\phi \approx\pm 8^\circ$ latitude\footnote{In this work, we use planetocentric latitude since the oblateness of Jupiter is small, and therefore, the difference between geocentric and geographic latitude is small.}, except for the jet at $20^\circ$ latitude, where $\Ro \approx 0.4$. Therefore, the flow is geostrophic to leading order. Additionally, variations of the Coriolis parameter $f$ over the flow scale $L$ are small: with $\beta = a^{-1} \partial_\phi f$ (Jupiter radius $a = 69.86\times10^{6}$ m, \citealp{Guillot99}), we have $\beta L/f < 0.2$ outside $\phi \approx \pm 8^\circ$ latitude. It follows that outside the equatorial jet (Fig.~\ref{f:obs}a), the leading-order geostrophic flow is approximately non-divergent, and vertical advection terms can be neglected in the zonal momentum balance. In a statistically steady state---a good approximation for Jupiter, as evidenced by the weak zonal-flow variations between the Voyager and Cassini observations \citep{Porco03,tollefson17}---the Eulerian mean zonal momentum equation thus is to leading order
\begin{equation}\label{e:zonmom}
f \Bar v \approx S - \Bar X.
\end{equation}
Zonal accelerations owing to molecular diffusion and/or smaller-scale turbulent Reynolds stresses are subsumed in $X$. If we neglect $\Bar X$---molecular diffusion is negligible, and there is no evidence smaller-scale turbulent Reynolds stresses are important at leading order in Jupiter's (or Earth's) free troposphere---the zonal momentum balance \eqref{e:zonmom} reduces to Eq.~\eqref{e:zonmom0} in the introduction. Because the mean meridional flow $\Bar v$ is entirely ageostrophic (the geostrophic meridional flow vanishes in the zonal mean), it is weaker by a factor $O(\Ro)$ than the approximately geostrophic mean zonal flow $\Bar u$. Hence, in midlatitudes (around $\pm 45^\circ$), where zonal flow speeds are of order $U\sim 10 \,\mathrm{m\,s^{-1}}$ but $\Ro \lesssim 0.02$, $\Bar v$ can at most be expected to be $O(0.2\,\mathrm{m\,s^{-1}})$.

The mean meridional flow $\Bar v$ on Jupiter is too weak to be inferred directly from observations \citep{Salyk06}. However, the horizontal eddy momentum flux divergence $S$ can be inferred by tracking observed cloud features. While there are quantitative uncertainties, its qualitative structure is unambiguous \citep{Salyk06}: at the level of the visible clouds, eddy momentum fluxes converge in prograde jets and diverge in retrograde jets (Fig.~\ref{f:obs}b). Using the observed $S$ (dashed blue line) to construct the implied mean meridional flow $\Bar v = S/f$ (orange line) reveals a weak flow, $O(0.03\,\mathrm{m\,s^{-1}})$, that converges in zones [$\partial_\phi (\Bar v\cos\phi) < 0$] and diverges in belts [$\partial_\phi (\Bar v\cos\phi) > 0$].

This pattern of meridional flow in the upper troposphere implies mean downwelling in zones and upwelling in belts in the lower troposphere. The eddy momentum fluxes driving this flow must have a strongly baroclinic structure and be confined to a shallow upper-tropospheric layer. Energetic constraints require this: if the observed eddy momentum fluxes extended deep into the atmosphere, the implied kinetic energy conversion rate would implausibly exceed the total energy supplied to the atmosphere by solar radiation and intrinsic heat combined \citep{Schneider09a,Liu10a}. Physically, this shallow structure may arises because wave activity (an invariant conserved by waves in the zonal mean) generated by baroclinic instability at lower levels propagates upward and turns meridionally as it encounters the strong static stability of the upper troposphere, creating the observed horizontal momentum fluxes \citep{Ait-Chaalal15a}. Because the driving fluxes and the \emph{meridional} branches of the circulation are shallow, but the circulation must close, its \emph{vertical} branches must extend deeply to connect to a layer where drag can act. This ``downward control'' principle \citep{Haynes91} links the upper-tropospheric dynamics to a deep drag layer, possibly of magnetohydrodynamic (MHD) origin \citep{Liu08,Schneider09a}. In contrast, alternative proposals of locally closing stacked cells driven by breaking gravity waves and their associated \emph{vertical} eddy momentum fluxes \citep{Ingersoll21} would require dynamics inconsistent with quasigeostrophic scaling and small Rossby numbers \citep[chapter~5]{Vallis06a}: under quasigeostrophic scaling, vertical eddy momentum fluxes are negligible in the zonal momentum balance (Eq. \ref{e:zonmom0}).

The conclusion appears inevitable, then, that to leading order, there is Eulerian mean downwelling in zones and upwelling in belts in Jupiter's lower troposphere. Assuming the upper branch of this circulation is distributed over a layer with a thickness comparable to the density scale height $H$, a rough estimate of the Eulerian mean vertical velocity below this layer is $\Bar w \sim (H/2) \, \mathrm{div}\, \Bar v = (H/2) \partial_\phi (\Bar v \cos\phi)/(a\cos\phi)$. This estimate for $H = 20 \,\mathrm{km}$ (appropriate for Jupiter's upper troposphere, where $T \sim 140 \, K$, $M \sim 2.2 \,\mathrm{g\,mol^{-1}}$ and $g \sim 26 \, \mathrm{m\,s^{-2}}$) is shown in Fig.~\ref{f:obs}d (dashed blue line). The vertical velocity is $O(10^{-4}\,\mathrm{m\,s^{-1}})$ and clearly is negative (downwelling) in zones and positive (upwelling) in belts. This apparently contradicts the inferences from the structure of clouds and hazes, from which it has been inferred that upwelling in zones leads to upper-tropospheric ammonia ice cloud formation and downwelling in the belts suppresses it. The apparent contradiction is resolved by considering an approximately Lagrangian mean flow.

\subsection{Transformed Eulerian mean flow}

\citet{Andrews76a,Andrews78a} showed that the Lagrangian mean flow advecting nearly conserved tracers can be approximated by the transformed Eulerian mean (TEM) flow. Other authors have also used TEM circulation to interpret Jupiter's atmospheric circulation, as discussed in \citet{Ingersoll17}, and to interpret the observed distribution of NH$_3$ below the weather layer in \citet{Duer21} and \citet{Lee21}. In a statistically steady state under quasigeostrophic scaling (which includes, in addition to the scaling assumptions of the previous section, the assumption that the thermal stratification is stable and relatively constant), the TEM flow has meridional and vertical components
\begin{equation}
\barstar{v}         = \Bar{v} - \frac{\partial}{\partial p}\left(\frac{\overline{v'\theta'}}{\partial_{p} \Bar \theta}\right) 
\label{e:vtar}
\end{equation}
and
\begin{equation}
\barstar{\omega} = \Bar{\omega} + \frac{1}{a\cos\phi}\, \frac{\partial}{\partial \phi} \left(\frac{\overline{v'\theta'}}{\partial_{p} \Bar \theta}\cos\phi\right).
\label{e:wtar}
\end{equation}
Here, $\omega = Dp/Dt$ is the vertical velocity in pressure coordinates, and $\theta$ is the potential temperature. The terms involving the meridional eddy flux of potential temperature $\overline{v'\theta'}$ represent approximately the Stokes drift. Adding the Stokes drift term on both sides of the Eulerian mean zonal momentum equation \eqref{e:zonmom0} gives the TEM zonal momentum equation,
\begin{equation}\label{e:TEM_mom}
f \barstar{v} = S -  f\frac{\partial}{\partial p}\left(\frac{\overline{v'\theta'}}{\partial_{p} \Bar \theta}\right) 
= - \frac{1}{a \cos\phi} \vec{\nabla}_{p}\cdot \vec{F}
\end{equation}
where we have written the right-hand side in terms of the divergence ($\vec{\nabla}_{p}\cdot{}$) of the Eliassen-Palm flux $\vec{F} = (F_{\phi}, F_{p})$ in the latitude-pressure plane \citep{Edmon80}, with 
\begin{equation}\label{e:EP_flux}
F_{\phi} = -\overline{u'v'}\, a \cos\phi \quad \text{and} \quad 
F_{p}    = f\frac{\overline{v'\theta'}}{\partial_{p}\Bar\theta}\, a \cos\phi.
\end{equation} 

The Eliassen-Palm flux is, under certain conditions, the flux of pseudomomentum (or wave activity), which is conserved by eddies (rather than by the total flow consisting of eddies and the Eulerian mean); hence, it is convenient in reasoning about how eddies interact with the mean flow and lead to mean transport of tracers \citep[e.g.,][]{Andrews87a}. The Eliassen-Palm flux is related to the eddy flux of quasigeostrophic potential vorticity,
\begin{equation}\label{e:qpv}
q  = f + \zeta + f\frac{\partial}{\partial p}\left(\frac{\theta}{\partial_{p} \Bar\theta}\right),
\end{equation}
with relative vorticity $\zeta$, by the Taylor-Bretherton identity \citep{Taylor15a,Bretherton66a,Edmon80}
\begin{equation}\label{e:taylor}
 \frac{1}{a \cos\phi} \vec{\nabla}_{p}\cdot \vec{F} =  \overline{v'q'}. 
\end{equation}
Combining the relations \eqref{e:TEM_mom} and \eqref{e:taylor} gives the well-known result that the TEM meridional flow is determined by the eddy flux of potential vorticity, which is conserved in adiabatic and frictionless eddy fluctuations \citep[e.g.,][]{Andrews87b}:
\begin{equation}\label{e:tempv}
f \barstar{v} = - \overline{v'q'}.
\end{equation}
The TEM vertical flow follows by continuity, $\vec{\nabla}_{p} \cdot (\barstar{v},~\barstar{\omega}) = 0$. These relations can be generalized to settings when quasigeostrophic scaling does not hold or when tracers are not nearly conserved  \citep[e.g.,][]{Andrews83,Tung86a,Andrews87b,Koh04,Schneider05a,Plumb05}. However, this is not necessary for a discussion of nearly conserved tracers in Jupiter's upper troposphere  (above $\about 400\,\mathrm{mbar}$), which is stably stratified and, for now, is our focus here.

The relation \eqref{e:tempv} allows us to derive a better approximation of the mean flow advecting tracers such as ammonia in Jupiter's upper troposphere. Unfortunately, the potential temperature flux contained in the potential vorticity or Eliassen-Palm flux \eqref{e:EP_flux} has not been directly measured. But as a first scaling estimate, we can assume the quasigeostrophic potential vorticity flux is downgradient and diffusive, 
\begin{equation}\label{e:diffus}
\overline{v'q'} \sim - K \partial_{y} \Bar q,
\end{equation} 
where $y=a\phi$ and $K>0$ is an eddy diffusivity---a standard approximation in geophysical fluid dynamics \citep[e.g.,][]{Rhines79b,Rhines82,Larichev95,Held96}. Unlike diffusive approximations for quantities such as angular momentum, which is not conserved by eddies, a diffusive approximation for quasigeostrophic potential vorticity is justifiable for two reasons  \citep{Corrsin74a,Rhines79b,Plumb79,Held00b}: (i) quasigeostrophic potential vorticity is only weakly non-conserved in the essentially frictionless and nearly adiabatic (rapid compared with radiative timescales) eddy fluctuations in Jupiter's upper troposphere; and (ii) the mean quasigeostrophic potential vorticity primarily varies on meridional scales that are large compared with eddy length scales \citep{Read06}. These two assumptions may not be satisfied accurately everywhere. For example, the mean quasigeostrophic potential vorticity has variations on the meridional scale of the zonal jets, which is similar to the eddy length scale (see below). But the diffusive approximation is a useful starting point for scaling arguments \citep[e.g.,][]{Held99b,Held99a,Schneider04a,Schneider06a}.

We assume the eddy diffusivity $K$ varies at most on scales that are large compared with meridional flow scales so that the diffusive approximation is justifiable. For the purpose of scaling arguments, we also neglect the modification of the diffusivity by the zonal jets themselves, which can inhibit meridional transport across their peaks \citep{Dritschel08,Nikurashin13a}. Then, the relation \eqref{e:tempv} with the diffusive closure \eqref{e:diffus} implies that the TEM meridional flow,
\begin{equation}\label{e:vbarstar_diffus}
\barstar{v} \sim \frac{K}{f} \partial_y \Bar q,
\end{equation} 
is strongly poleward where the quasigeostrophic potential vorticity gradient $\partial_{y} \Bar q$ is strongly positive; it is more weakly poleward or equatorward where $\partial_{y} \Bar q$ is more weakly positive or negative. Now, the quasigeostrophic potential vorticity gradient $\partial_y \Bar q$ in Jupiter's upper troposphere at least above $\about 270\,\mathrm{mbar}$ is dominated by the barotropic component $\partial_y \Bar q \approx  \beta + \partial_y \Bar \zeta$, the gradient of the absolute vorticity $f+\Bar\zeta$; the stretching term only modifies $\partial_y \Bar q$ by $\about 10\%$ \citep{Read06}. Hence, we ignore the stretching term here, although its contribution is more likely to be significant at lower levels, where the thermal stratification is close to neutral. The absolute vorticity gradient in the upper troposphere is locally enhanced in the cores of prograde jets (where $\partial_y \Bar \zeta \approx -\partial_{yy} \Bar u > 0$ because the zonal flow curvature is negative; it is reduced in the cores of retrograde jets (where $\partial_y \Bar \zeta < 0$ because the zonal flow curvature is positive). So there is a correlation between zonal wind extrema and the absolute vorticity gradient, as also seen in the analysis of \citet{Read06}; however, this does not mean that the zonal wind and absolute vorticity gradient are strictly proportional, as an analysis of a 1.5-layer model with assumed barotropic deep wind structures has suggested \citep{Dowling93}.  The planetary vorticity gradient $\beta$ varies only on scales larger than the meridional jet spacing and hence does not contribute substantially to modulations of the absolute vorticity gradient on the scales of zones and belts (Fig.~\ref{f:obs}c). It follows that the TEM meridional flow $\barstar{v}$ can be expected to be relatively strongly poleward in prograde jets and more weakly poleward or even equatorward in retrograde jets. This leads to convergence in belts and divergence in zones. It resembles the ad hoc suggestion by \citet{Gierasch86a} for a TEM meridional flow that alternates between prograde and retrograde jets. The distinction between that work and our work is that the TEM meridional flow we suggest does not need to change sign between jets but may be predominantly poleward throughout the upper troposphere, only with strength variations between jets.

To be concrete, let us assume the diffusivity is constant and, as is standard in turbulence theory \citep[e.g.,][]{Wyngaard10a} generally and in quasigeostrophic turbulence theory \citep[e.g.,][]{Held96} specifically, can be estimated from the meridional eddy velocity scale $V \sim 3\,\mathrm{m\,s^{-1}}$ \citep{Salyk06} and the meridional eddy length $L\sim 2000\,\mathrm{km}$ (which is of the order of the Rossby deformation radius, e.g., \citealp{Cho2001, Li06, Thomson16a, Young17a}). Then, $K \sim VL \sim 6 \times 10^6 \,\mathrm{m\,s^{-2}}$. Figure~\ref{f:obs}c shows the TEM meridional flow (orange line) that results when the quasigeostrophic potential vorticity gradient $\partial_{y} \Bar q$ is approximated by the absolute vorticity gradient $\partial_{y}(f + \Bar\zeta)$ (dashed blue line) obtained from the observed zonal flow in Fig.~\ref{f:obs}a \citep{Salyk06}. The resulting TEM meridional flow is of order $\barstar{v} \sim K\beta/f \sim \mathrm{0.3\,m\,s^{-1}}$ in midlatitudes---an order of magnitude stronger than the Eulerian mean meridional flow. As the absolute vorticity gradient is generally positive, except in some retrograde jets \citep{Ingersoll81,Salyk06}, the TEM meridional flow likewise is generally poleward. It is clear from Fig.~\ref{f:obs}d that the TEM meridional flow implied by the diffusive potential vorticity flux closure converges in belts [$\partial_\phi (\barstar{v}\cos\phi) < 0$] and diverges in zones [$\partial_\phi (\barstar{v}\cos\phi) > 0$]. Because the TEM mass circulation almost certainly closes at depth for reasons analogous to those for the Eulerian mean circulation (with potential vorticity fluxes replacing angular momentum fluxes in the arguments), this suggests that in the underlying lower troposphere, there is TEM downwelling in belts and upwelling in zones. The rough estimate $\barstar{w} \sim (H/2) \, \mathrm{div}\,\barstar{v}$ of the TEM vertical velocity below the upper branch of the TEM mass circulation is shown in Fig.~\ref{f:obs}d (orange line). This TEM vertical velocity is of order $O(10^{-3}\,\mathrm{m\,s^{-1}})$---an order of magnitude stronger than the corresponding Eulerian mean vertical velocity. It clearly is positive (upwelling) in zones and negative (downwelling) in belts. This TEM vertical velocity is about an order of magnitude stronger than the TEM vertical velocity around $\about 100$~mbar  estimated from radiative transfer calculations and observed properties of Jupiter's upper atmosphere \citep{West92a,Moreno97a}. There is no contradiction, however, as our estimates apply to the layer of the visible cloud tops (below 270~mbar); the TEM vertical velocity can be expected to decrease upward toward the tropopause (at $\about 100$~mbar).

Thus, the TEM circulation inferred by the diffusive potential vorticity flux closure is much stronger and has the opposite direction of vertical motion than the Eulerian mean circulation. It is qualitatively consistent with the observed structure of clouds and hazes (Fig.~\ref{f:obs}d). Based on our scaling arguments, barotropic variations of the quasigeostrophic potential vorticity gradient are responsible for the jet-to-jet variations of potential vorticity fluxes. These meridionally varying potential vorticity fluxes, in turn, are associated with a TEM meridional flow that is more strongly poleward in prograde jets and less strongly poleward or even equatorward in retrograde jets, and so, at deeper levels, drives upwelling in zones and downwelling in belts. While the meridional flow in the Eulerian mean circulation is associated with eddy momentum flux divergence $S$, or with the barotropic component of the potential vorticity flux, the much greater strength of the TEM circulation implies that the baroclinic component of the potential vorticity flux, involving the meridional heat flux in Eq.~\eqref{e:TEM_mom},  dominates. The baroclinic component of the potential vorticity flux is associated with an eddy mass flux along isentropes \citep[][chapter~7.3]{Schneider05a,Vallis06a}, which represents a Stokes drift and accomplishes the TEM circulation. Note that there is no contradiction in assuming that the \emph{mean} potential vorticity gradient in the upper troposphere is dominated by its barotropic component, whereas the \emph{eddy} flux of potential vorticity is predominantly baroclinic.

\begin{figure*}[htb]
\centerline{\includegraphics[width=1.5\columnwidth]{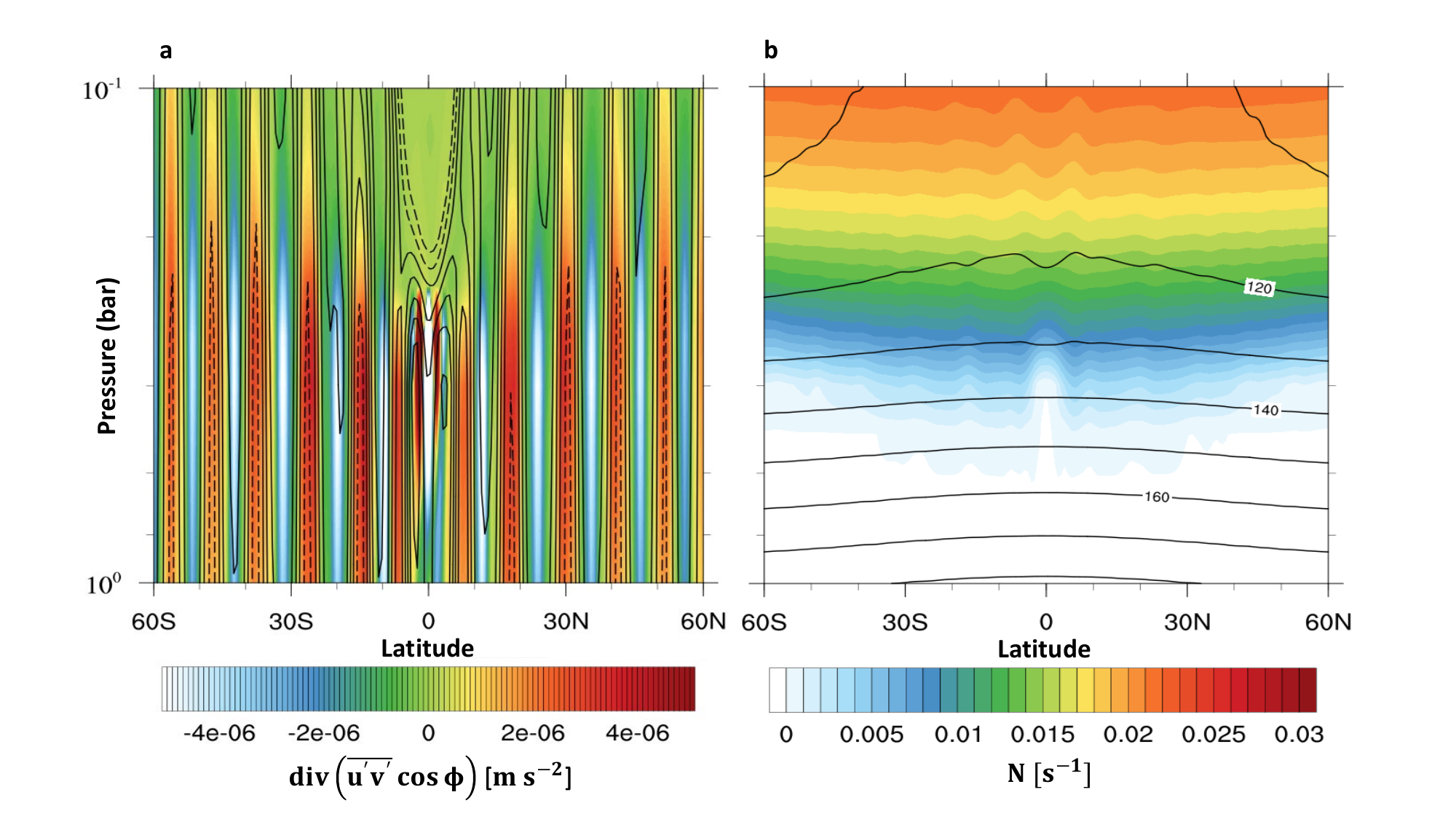}}
\caption{Mean flow fields in the latitude-pressure plane in the upper atmosphere of the simulation above 1~bar from \citet{young19a,young19b}. (a) Zonal velocity (line contours) and horizontal eddy momentum flux divergence $S$ (colours). The line contours represent zonal velocities between $-8$ and $28 \, \mathrm{m \, s^{-1}}$, with a contour interval of $5 \, \mathrm{m \, s^{-1}}$. Solid contours are for prograde flow, and dashed contours are for retrograde flow.  (b) Temperature (contours, contour interval 10~K) and buoyancy frequency $N$ (colours).}
\label{f:sim3D}
\end{figure*}

\section{Demonstration with global circulation model}
\label{sec:gcm}
We demonstrate the validity of the scaling arguments with simulation results from the global circulation model (GCM) of Jupiter's upper atmosphere that was developed and presented in \citet{young19a,young19b}\footnote{The GCM data used in this work can be found in \cite{young2018data}:
10.5287/bodleian:PyYbbxpk2.}. The Jupiter GCM is based on the finite-volume MITgcm \citep{marshall97a,marshall97b,adcroft04}, which solves the primitive equations on a spherical latitude-longitude grid. The GCM has an artificial lower boundary at 18~bar, which is necessary to make it computationally feasible to resolve large-scale eddies in the upper atmosphere while remaining consistent with Jupiter's observed energetics \citep{young19a}. At the lower boundary, a linear drag mimics, in a simplified way, the MHD drag on the flow that occurs in Jupiter at much greater depth \citep{Liu08,Liu13a}. Additionally, a spatially uniform upwelling heat flux representing Jupiter's intrinsic heat flux is imposed. A semi-grey radiative transfer scheme, similar to that used in \citet{Schneider09a}, is utilized to represent radiative processes. To improve accuracy in the deepest layers with large optical depths, the radiation scheme utilizes an exponential dependence of the Planck function on optical depth for the thermal infrared radiation part. This approach was discussed in \citet{fu93} and \citet{mendonca15}. The absorbed solar radiative flux is $\mathrm{8.2\,W\,m^{-2}}$ in the global mean, and the imposed intrinsic heat flux at the lower boundary amounts to $\mathrm{5.7 \, W\,m^{-2}}$ \citep{read2016}. To study the layered cloud structures in Jupiter \citep{Zuchowski09b, young19b}, the GCM allows the passive advection of seven tracers: ice and gaseous ammonia (NH$_3$),  gaseous, liquid and ice water (H$_2$O), gaseous hydrogen sulfilde (H$_2$S), and ice ammonium hydrosulfide (NH$_4$SH). The GCM uses Jupiter's planetary radius, rotation rates, and other physical parameters. Further details can be found in \citet{young19a,young19b}. 

The results from the Jupiter GCM were obtained using a horizontal spatial resolution of 0.7$^\circ$, which is necessary to resolve the first baroclinic Rossby radius and to simulate Jupiter's meteorology \citep{Achterberg89,Vasavada05,Read06}. The simulations were integrated over 130,000–150,000 Earth days. Long-time integration is required due to the slow thermal and dynamical adjustment of the deep atmosphere. 

The GCM reproduces qualitatively many of the observed features of Jupiter's flow and temperature structure. For example, in the upper troposphere, it reproduces a prograde equatorial jet and weaker alternating retrograde and prograde off-equatorial jets (Fig.~\ref{f:sim3D}a, contours). Eddy momentum fluxes converge in prograde jets and diverge in retrograde jets (Fig.~\ref{f:sim3D}a, colors), with $S$ reaching $O(10^{-5}\,\mathrm{m\,s^{-2}})$ in the upper troposphere---as observed on Jupiter (cf.~Fig.~\ref{f:obs}b). As in the simulation, it must also be the case on Jupiter that the strongest eddy momentum flux convergence/divergence is confined to the upper troposphere because otherwise the kinetic energy transfer from eddies to the mean flow would exceed the energy available to drive the flow from the absorbed insolation and intrinsic heat fluxes \citep{Schneider09a,Liu10a}. The simulated eddy momentum flux convergence and divergence pattern is consistent with baroclinic eddy generation that occurs preferentially in the baroclinically more unstable prograde jets \citep{Schneider09a,Liu10a,Liu11a,young19a}. In a previous 3D model by \citet{Lian10}, moist convective storms were shown to drive the formation of a zonally banded circulation. However, this required energy fluxes at the lower boundary that were stronger than the observed intrinsic heat fluxes. This limitation was later addressed in \citet{Lian23}. The model from \citet{young19a,young19b} does not include moist convection, which may affect the thermal stratification but is unlikely to directly affect the large-scale circulation \citep{Emanuel94a}. These results are supported by the difference in scales between the typical sizes of individual moist convective storms and the estimated scale of energy injection in the observed turbulent kinetic energy cascade \citep{Young17a}. The observed scales suggest that baroclinic eddies are more likely than convective storms to be the primary sources of energy for large-scale flows. Additionally, the GCM of \citet{young19a,young19b} reproduces the thermal structure of Jupiter's troposphere, including a neutrally stratified lower troposphere overlaid by a stably stratified upper troposphere that is capped by a tropopause (Fig.~\ref{f:sim3D}b). 

Thus, while this simulation does not replicate Jupiter's upper atmosphere exactly, it does reproduce its central dynamical balances as far as they are observed, and it can serve as a testbed to investigate the validity of the theoretical arguments we presented.

\subsection{Upper-tropospheric dynamics}

\begin{figure}[]
\centerline{\includegraphics[width=0.7\columnwidth]{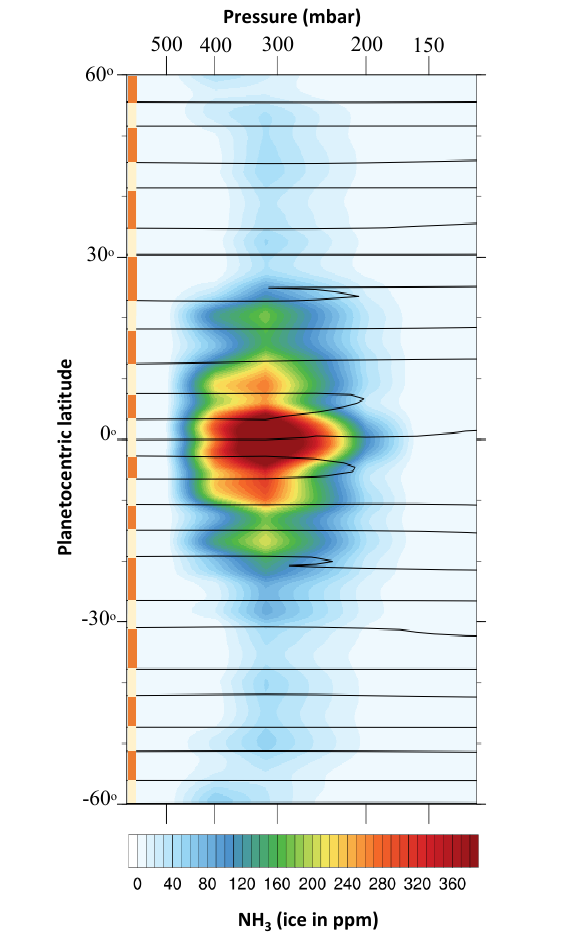}}
\caption{Latitude-pressure contour map representing the mass mixing ratio of ice NH$_3$ in Jupiter's upper troposphere \citet{young19a,young19b}. The GCM generates distinct zonal band patterns, which are closely linked to the pattern of the zonal winds. Belts are bands with low cloud density and cyclonic vorticity, while zones have higher cloud density and anticyclonic vorticity. The solid black lines represent regions where the vorticity changes sign and the edges of the zonal band patterns. The orange rectangles indicate the latitudinal extent of the belts. These are areas of cyclonic shear zones in the simulated atmospheric flow, following the same definition as in Fig.~\ref{f:obs}, which used observational data.}
\label{f:vr_cld}
\end{figure}

The Jupiter GCM produces a cloud band structure that resembles the one observed. For a detailed comparison of the model results with observations of Jupiter, we refer to \citet{young19b}. Figure~\ref{f:vr_cld} shows the concentration of ice NH$_3$ in the simulation, along with the belts and zones as identified by the relative vorticity field. As observed, higher concentrations of ice NH$_3$ occur in zones with anticyclonic meridional shear of the zonal wind, while lower concentrations of ice NH$_3$ are found in bands with cyclonic shear \citep{young19b}. The two hemispheres have a similar band structure, although they are not completely symmetrical. As observed, prograde jets are located on the equatorward flanks of the belts and retrograde jets on the poleward flanks. Each zonal structure is defined by its edges, which are located in regions where the vorticity changes sign. The more clearly defined zonal structures of Jupiter's observed belts and zones indicate overly diffusive dynamics of the underresolved simulation.

Interestingly, as shown in \citet{young19b}, the Jupiter simulations used here also produce an equatorial plume of NH$_3$ vapor similar to the one observed by Juno \citep{li17}. However, as explained in \citet{young19b}, as the simulation progresses, more vapor is lifted in the tropical regions and spreads to higher latitudes due to turbulent mixing. In mid-latitudes, meridional circulation cells also slowly lift NH$_3$ vapor from the model's deeper layers, broadening the initial NH$_3$ plume latitudinally. The initial narrow equatorial plume produced in the simulation is a transient phenomenon that emphasizes the role of the upwelling in the equatorial region \citep{young19b}. 

\subsubsection{Mean meridional circulation}

\begin{figure}[]
\centerline{\includegraphics[width=1.3\columnwidth]{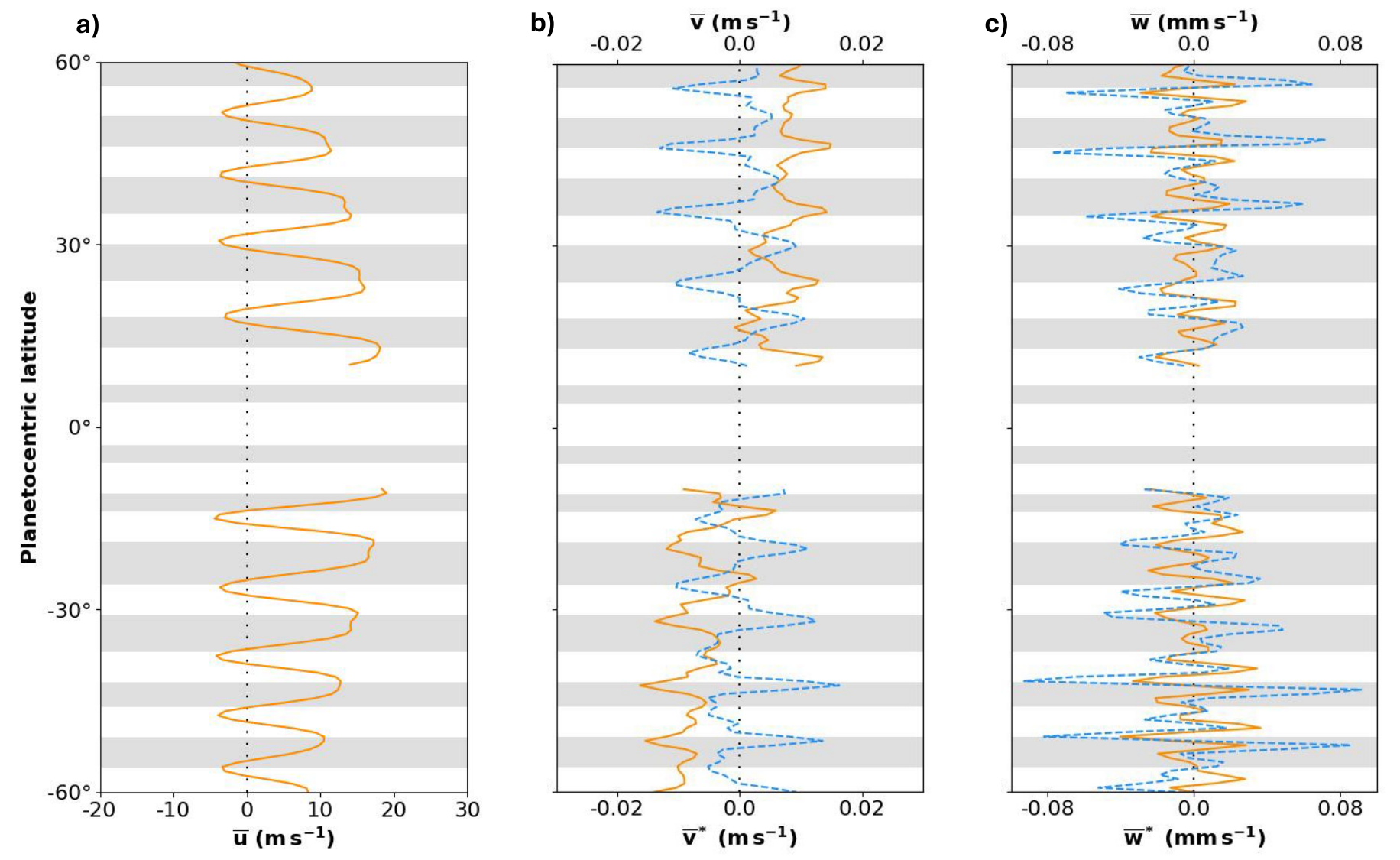}}
\caption{Simulated zonal (left panel), meridional (middle panel) and vertical (right panel) velocities as a function of latitude (degrees) at Jupiter's GCM NH$_3$ cloud top (200~mbar). Grey-shaded regions mark belts (cyclonic shear zones), as in Fig.~\ref{f:obs}. The middle panel shows the Eulerian mean meridional velocity (dashed blue line) and the transformed Eulerian mean meridional velocity (orange line). Both meridional velocities are in units of $\mathrm{m~s^{-1}}$. The right panel shows the Eulerian mean vertical velocity (dashed blue line) and the transformed Eulerian mean vertical velocity (orange line). Note that the vertical velocities in the GCM are in Pa$\,$s$^{-1}$ and were converted to mm$\,$s$^{-1}$ using the hydrostatic relation.}
\label{f:v_w}
\end{figure}

To determine the validity of our theoretical arguments in the previous sections, we plotted the meridional and vertical components of the Eulerian and transformed Eulerian mean circulations in Fig.~\ref{f:v_w}b and \ref{f:v_w}c. To make it compatible with the observations, the values plotted correspond to the upper cloud region in the GCM simulations (Fig.~\ref{f:vr_cld}), above the weakly stratified region (Fig.~\ref{f:sim3D}b). The results correspond to a pressure level of approximately 200~mbar.\footnote{Note that the simulated atmosphere's weak static stability compromises the standard TEM flow's accuracy (Eqs.~\ref{e:vtar} and \ref{e:wtar}) for depths greater than 300~mbar. At the 300~mbar pressure level, the impact of static stability on TEM flow becomes evident as $\barstar{v}$ increases rapidly towards higher latitudes, where static stability decreases quasi-monotonically with latitude.} The Eulerian mean meridional flow (black lines) exhibits a similar pattern to that inferred from observations in Fig.~\ref{f:obs}: convergence in zones and divergence in belts. By contrast, and in agreement with the theoretical arguments presented earlier, the TEM meridional flow (red line) generally exhibits convergence in belts and divergence in zones. However, the TEM meridional flow is of similar magnitude as the Eulerian mean flow component, unlike what we inferred from observations (Fig.~\ref{f:obs}b). The main differences between the simulations and observations lie in the magnitude of the meridional eddy velocity and the horizontal eddy momentum flux divergence $S$.  The meridional eddy velocity has a magnitude in the simulations of around $\sim 1\,\mathrm{m\,s^{-1}}$, which is a factor of 3 smaller than our estimates based on the observations by \cite{Salyk06}. Observations also show a stronger eddy momentum flux divergence compared with the simulations at the 200-mbar pressure level. Fig.~\ref{f:sim3D}a shows a decrease in the horizontal eddy momentum flux divergence as the altitude increases from around the 400-mbar pressure level. However, the TEM flow is not computed for pressure levels deeper than 200~mbar because of the low static stability at deeper levels. Nonetheless, the results emphasize the importance of the horizontal eddy momentum flux divergence and meridional eddy velocity in interpreting TEM flow components, which is consistent between simulations and observations.

For the TEM vertical flow, it is important to highlight that the GCM simulations show, on average, a pattern that aligns with the scaling arguments from the previous section. Our results help to resolve the inconsistency in the contrasts of belt-zone vertical motions and cloud structure in Jupiter's upper troposphere \citep{fletcher20}. The pattern of the TEM vertical flow is consistent with the convergence and divergence regions of the TEM meridional flow, with enhanced upwelling in zones (Fig.~\ref{f:v_w}). The consistency with the scaling arguments and observations is stronger at the center of the bands. However, it becomes more complex near the band's edges, which show larger vertical wind shears. These wind shears might contribute to turbulent mixing between different band structures. This mixing is more pronounced in the simulations than in the observations, resulting in less confined belts/zones in the simulations compared to the observations. The magnitudes of the vertical flow in the TEM and Eulerian mean are similar in the simulation but weaker than the values inferred from observations. Such a trend is also seen in the meridional direction analysis, as described above. 

The standard TEM formulation explored above becomes poorly defined for deeper pressure levels as the method requires division by static stability, which decreases rapidly below the pressure level plotted in Fig. \ref{f:v_w}. To study the extension of the atmospheric cells with depth and overcome this limitation, we apply a modified residual circulation from \citet{Held99a}, which does not require division by static stability. The selection of the direction in the eddy flux component of the residual circulation is arbitrary \citep{Andrews78a}, and following \citet{Held99a}, we can replace the meridional eddy heat flux with the vertical component to redefine the residual circulation in regions of low static stability. The modified residual circulation can be defined in terms of the mixing slope $S$ and the slope of the mean isentropes $I$ as follows \citep{Held99a}:
\begin{equation}
S \equiv \frac{\overline{\omega'\theta'}}{\overline{v'\theta'}},
\label{e:S_eq}
\end{equation}
\begin{equation}
I \equiv -\frac{1}{a\cos\phi}\frac{\partial_\phi\Bar{\theta}\cos\phi}{\partial_p\Bar{\theta}},
\label{e:isentropic_slope}
\end{equation}
\begin{equation}
\barstar{\omega}_\dagger = \Bar{\omega} + \frac{1}{a\cos\phi}\, \frac{\partial}{\partial \phi} \left[\left(\frac{S}{I}\right)\frac{\overline{v'\theta'}}{\partial_{p} \Bar \theta}\cos\phi\right].
\label{e:wstar_modified}
\end{equation}
If $S/I < 0.05$, we use the conventional expression defined in Eq. (\ref{e:wtar}); we use the modified expression Eq. (\ref{e:wstar_modified}) if $S/I > 0.05$, where the terms related to the static stability cancel out and the vertical, rather than meridional, eddy heat flux appears in the residual velocity. The threshold of $0.05$ was chosen to allow the solutions at approximately 200~mbar pressure level to be all calculated using the standard TEM formulation and being consistent with Fig. \ref{f:v_w}, while maintaining a smooth transition to the modified TEM formulation. In the upper atmosphere, the modified TEM is, in general, smaller than the traditional TEM. See \citet{Held99a} for more details and interpretation of this modified equation.

\begin{figure*}[htb]
\centerline{\includegraphics[width=1.35\columnwidth]{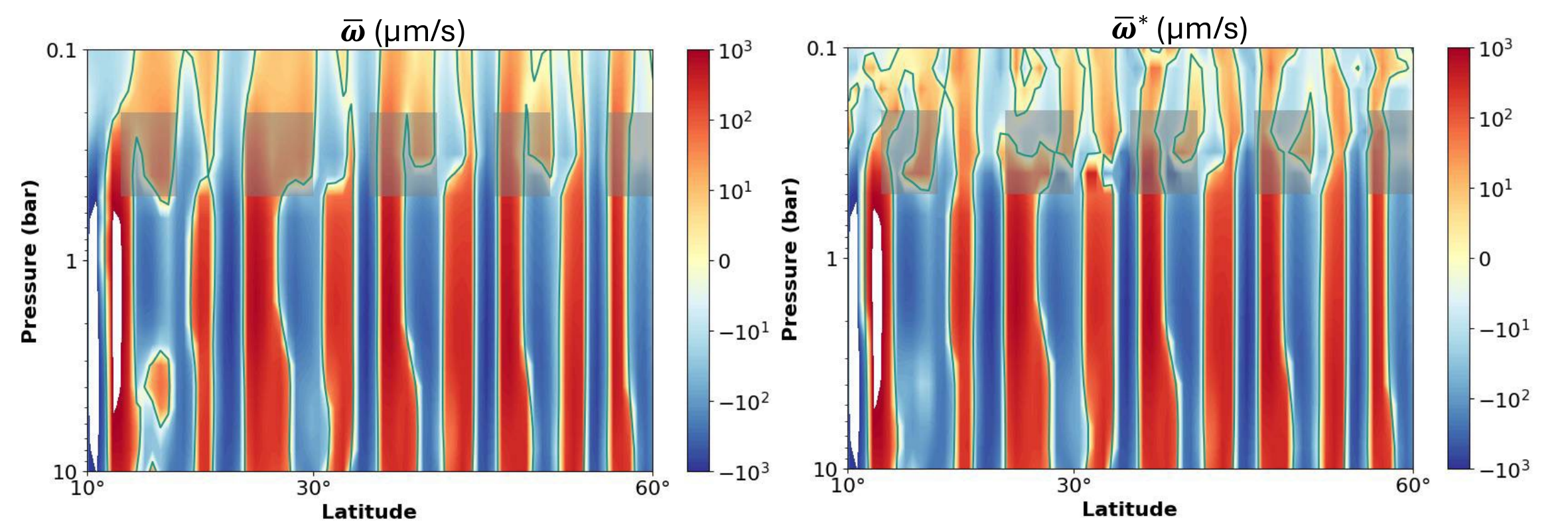}}
\caption{The Eulerian and transformed Eulerian mean vertical flow fields in the latitude-pressure plane calculated from the GCM simulation of \citet{young19a,young19b}. In both panels, the belt regions are marked with grey rectangles. In the left panel is the Eulerian mean vertical velocity, which shows upper-tropospheric upwelling in the belts and downwelling in the zones. The transformed Eulerian mean vertical velocities, shown on the right panel, have the main differences compared to the left panel in the upper troposphere, where, at least at the center of each band, downwelling predominates in belts and upwelling in zones.  Below the 200~mbar level, the transformed Eulerian mean closely resembles the Eulerian mean, primarily because the eddy-induced term on the right side of Eq.~\ref{e:wstar_modified} is, in general, smaller than the Eulerian term. This figure illustrates the atmospheric cells that extend from the upper troposphere down to the bottom of the model domain, where they close in a frictional layer, consistent with the principle of ``downward control'' of atmospheric overturning circulations \citep{Haynes91}. Note that, as in Fig. \ref{f:v_w}, we converted the vertical velocities from Pa$\,$s$^{-1}$ to mm$\,$s$^{-1}$ using the hydrostatic relation.}
\label{f:wstar_2D}
\end{figure*}

Fig.~\ref{f:wstar_2D} illustrates the vertical extension of the Eulerian mean vertical velocity and the residual vertical flow, derived from combining Eqs. (\ref{e:wtar}) and (\ref{e:wstar_modified}). This figure aligns with the results presented in Fig. \ref{f:v_w}, where the residual winds at the center of the band structure in the upper troposphere exhibit predominantly downwelling in belts and upwelling in zones, in contrast to the Eulerian mean. Additionally, this figure demonstrates the ``downward control" principle of atmospheric overturning circulations \citep{Haynes91}: the atmospheric cells extend from the upper troposphere to the bottom of the model domain, where they ultimately close in a frictional Ekman layer at depth. As previously mentioned, large complex structures also form near the edges of the bands, likely due to turbulent mixing between different band structures. However, the magnitude of this complex band structure near the edges may be influenced by the inadequate representation of sub-grid turbulent mixing in the model compared to Jupiter's atmospheric conditions.

\section{Conclusions}

The angular momentum balance provides powerful constraints based on already available observations. In this work, we used fundamental principles of the angular momentum balance, together with simple diffusive closures for potential vorticity fluxes, to physically constrain the atmospheric circulation in Jupiter's upper troposphere and reasoned how it extends to deeper levels. We resolved the apparent contradiction between the contrasts of belts and zones and vertical motion in Jupiter's upper troposphere by considering the transformed Eulerian mean (TEM) circulation to represent the Lagrangian mean transport of tracers. We take into account the contribution from turbulent eddies to the tracer transport that represents the Stokes drift in the upper cloud region, where static stability is larger compared to the lower troposphere and away from the equator (latitudes between 10 and 60 degrees in both hemispheres). Based on the zonal momentum balance with diffusive closures for potential vorticity fluxes, our scaling arguments show TEM upward vertical motion in zones and downward motion in belts, resolving the apparent contradiction with the Eulerian mean flow found in observations. 

The observed circulation is generated from the top down of the atmosphere in a ``downward control'' fashion, as also happens on Earth \citep{Haynes91}. In Jupiter, the circulation is driven by eddy fluxes generated in the upper troposphere and closed in the deep atmosphere, likely in an Ekman layer with magnetohydrodynamic (MHD) drag. Our proposed circulation in Jupiter is consistent with the deep jet stream distribution inferred from Juno spacecraft observations \citep[e.g.,][]{Kaspi18a}. While we need to displace the MHD drag layer to shallower levels in the GCM for computational reasons (making it deeper leads to exponentially longer equilibration times), doing so still leads to a physically consistent closure of the circulation, with the Eulerian circulation extending downward along surfaces of constant specific angular momentum (Fig.~\ref{f:wstar_2D}). 

Our results still do not fully explain the global distribution of ammonia (NH$_3$) revealed by the Juno mission (e.g., \citealt{bolton17}; \citealt{li17}), requiring further investigation. To gain a comprehensive understanding of the distribution of ammonia (NH$_3$) in Jupiter's atmosphere, it is necessary to consider the impact of various physical processes, such as dynamical transport, sources (chemical products, evaporation of precipitates), and sinks (photochemical destruction, condensation). Observations from Juno \citep[e.g.,][]{li17} show a plume of NH$_3$ in the tropics adjacent to bands of NH$_3$-poor air that appears to be sinking. From the maps of the NH$_3$ mass distribution, \citet{Ingersoll17} argue that there is a net upward transport of ammonia in Jupiter that needs to be resolved since there is no NH$_3$ rain or chemical reactions to close the  NH$_3$ budget. \citet{Ingersoll17} estimate that spherical NH$_3$ droplets with diameters 1-5 mm evaporate before reaching pressures deeper than 1-1.5~bar. Juno's microwave radiometer experiment (MWR) observed a change in microwave brightness gradient with depth. In \cite{fletcher21}, it was found that belts change from being depleted to being enriched in NH$_3$ as a function of pressure. Levels with pressure values lower than 5~bar are depleted in NH$_3$ while levels with pressure values higher than 10~bar are enriched in NH$_3$. We have used the results presented in this work to infer the Eulerian mean and TEM circulation in the deep atmosphere; more complex models are needed to represent the physics in the deep atmosphere robustly and better elucidate the main mechanisms driving the global ammonia distribution. Below the weather layer, \citet{Lee21} and \citet{Duer21} have utilized a TEM scaling analysis to examine atmospheric dynamics based on the observed ammonia distribution \citep{bolton17}. The results from these two studies suggest the formation of multiple Ferrel-like cells, characterized by upward movement in the belts and downward motion in the zones. While these studies were able to explain some of the observed patterns in the distribution of ammonia in the deep atmosphere, they based their models on the assumption of a multi-tier circulation \citep{Ingersoll00,Showman05a}. This complex circulation has also been proposed to explain the preferred conditions for lightning formation within the belts \citep{Little99a,Gierasch00,Porco03}. However, it is important to note that thunderstorms can occur in areas with mean subsidence, for example, in Earth's subtropics and lower midlatitudes  (e.g., \citealt{Christian03a}, \citealt{Houze2004}). Additionally, no robust physical mechanism has been proposed to close the circulation near the ammonia clouds. This would require a rapid change in the eddy momentum flux divergence in the region with over-lapping vertical cells, for which there is neither observational evidence nor a convincing physical mechanism. To address the circulation near the ammonia cloud region, \citet{Ingersoll21}  proposed that a two-tiered circulation could be driven by small-scale gravity waves that propagate both upward and downward from the ammonia cloud layer. The vertical momentum flux associated with wave breaking can help close the momentum budgets. However, as we noted previously, the vertical eddy momentum fluxes are negligible in the zonal momentum balance (see Eq.~\ref{e:zonmom0}) when the Rossby number is small and the stratification is stable, according to quasigeostrophic scaling \citep[chapter~5]{Vallis06a}. Consequently, closing the circulation using vertical eddy momentum fluxes would require dynamics at larger Rossby numbers, which is inconsistent with Jupiter's observed dynamical regime outside of the equatorial region.

Our proposed TEM circulation with enhanced divergence in zones and convergence (or reduced divergence) in belts is most naturally formulated outside the equatorial zone, where the Rossby number is small (on Jupiter, $\mathrm{Ro} \lesssim 0.1$ for eddies outside about $10^\circ$ latitude). Our theoretical framework offers an explanation for the observations from Voyager IRIS \citep{Carlson94a}, which found that belts are warmer than zones. Above the cloud layer, the warmer zones imply a change in the temperature gradient in the upper troposphere \citep{fletcher20,Simon-Miller06}. The thermal-wind relation then implies that zonal jets weaken with height, and indeed, IRIS/CIRS thermal-wind analyses off the equator show zonal jets decaying in the upper troposphere (e.g., \citealt{Gierasch86a}). However, we emphasize that this evidence is strongest in Jupiter’s tropical belt–zone system. In mid-latitudes, the contrasts between belts and zones (such as in visible color and temperature) are significantly weaker and more variable compared to those in the equatorial region (e.g., \citealt{Antunano20,fletcher20}), and it is less clear to what extent they are consistent with thermal inferences from the TEM framework.


Our proposed theoretical arguments have been validated through 3D GCM simulations of Jupiter. These simulations qualitatively capture the main dynamical properties in Jupiter's troposphere. Similar to what is seen in observations, regions with higher concentrations of ice ammonia (zones) experience upward vertical motion, while regions with lower concentrations of ice ammonia (belts) experience downward vertical motion.

The circulation proposed in this study aligns with previous works by \citet{Schneider09a} and \citet{Liu10a}, emphasizing the importance of using complex 3D models to interpret observational data and robust physical arguments that, in particular, close the angular momentum balance to understand Jupiter's atmospheric circulation.

\paragraph{Acknowledgments}
This research was supported by the NASA Outer Planets Research Program (Grant NNX10AQ05G), a David and Lucile Packard Fellowship, the Carlsberg Research Stays Grant (CF22-0011) and the UKRI Horizon Europe Guarantee grant (EP/Z00330X/1). We thank Colette Salyk for providing the zonal velocities and eddy momentum fluxes inferred from Cassini's observations.




\bibliographystyle{model2-names}

\end{document}